\documentclass[aps,prl,twocolumn,showpacs,groupedaddress]{revtex4}
\usepackage{amsmath,amsfonts,bm}
\usepackage{graphicx,epsfig}
\usepackage{color}

\usepackage{epstopdf}

\begin{document}

\title{Enhanced Microfluidic Mixing via a Tricritical Spiral Vortex Instability}

\author{Simon J. Haward,$^{1}$ Robert J. Poole,$^{2}$ Manuel A. Alves,$^{3}$
Paulo J. Oliveira,$^{4}$ Nigel Goldenfeld,$^{5}$}

\author{Amy Q. Shen$^{1}$}

\affiliation{$^{1}$Okinawa Institute of Science and Technology Graduate University,
Onna, Okinawa 904-0495, Japan}

\affiliation{$^{2}$School of Engineering, University of Liverpool$\mathrm{,}\thinspace$Brownlow
Street, Liverpool L69 3GH, United Kingdom}

\affiliation{$^{3}$Faculdade de Engenharia da Universidade do Porto$\mathrm{,}\thinspace$Centro
de Estudos de Fen\'{o}menos de Transporte, Rua Dr. Roberto Frias, 4200-465
Porto, Portugal}

\affiliation{$^{4}$Departamento de Engenharia Electromec\^{a}nica, C-MAST$\mathrm{,}\thinspace$Universidade
Beira Interior, 6201-001 Covilh\~{a}, Portugal}

\affiliation{$^{5}$Department of Physics$\mathrm{,}\thinspace$University of
Illinois at Urbana-Champaign, Loomis Laboratory of Physics, 1110 West
Green Street, Urbana, Illinois 61801-3080}

\begin{abstract}
Experimental measurements and numerical simulations are made on fluid flow through cross-slot devices with a range of aspect (depth:width) ratios, $0.4\leq\alpha\leq3.87$. For low Reynolds numbers Re, the flow is symmetric and a sharp boundary exists between fluid streams entering the cross-slot from opposite directions. Above
an $\alpha$-dependent critical value $20\lesssim\textnormal{R}\textnormal{e}_{c}(\alpha)\lesssim100$, the flow undergoes a symmetry-breaking bifurcation (though remains steady and laminar) and a spiral vortex structure develops about the central axis of the outflow channel. An order parameter characterizing the instability grows according to a sixth-order Landau potential, and shows a progression from second order to first order transitions as $\alpha$ increases. A tricritical point occurs for $\alpha\approx0.55$. The spiral vortex acts as a mixing region in the flow field and this phenomenon can be used to drive enhanced mixing in microfluidic devices.
 
\end{abstract}

\pacs{05.70.Fh, 47.20.Ky, 47.61.Ne, 64.60.Kw}

\maketitle

The ability of fluids to mix is greatly enhanced by turbulence, which occurs at large values of the Reynolds number $\textnormal{Re}\equiv UL/\nu$, where $U$ and $L$ are characteristic velocity and length scales respectively and $\nu$ is the kinematic viscosity of the fluid. Small length scales tend to suppress Re, making it difficult to develop turbulent mixing in microfluidic devices. Achieving efficient mixing via diffusion dominated processes presents a major technological challenge to the expanding field of lab-on-a-chip development.

\begin{figure}
\includegraphics[width=3.3in]{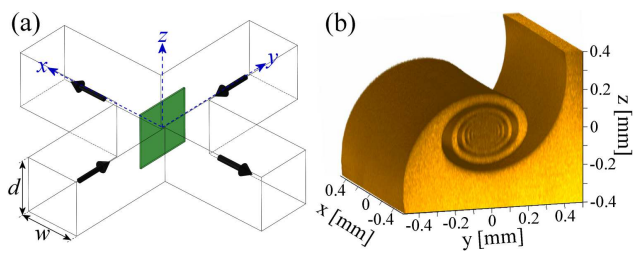}
\caption{(Color online) (a) Schematic diagram of a cross-slot device. Flow
enters along $y$ and exits along $x$. Confocal microscopy is performed
in $z$-planes, which are scanned through the full depth of the device
and used to reconstruct an image in the $x=0$ plane (shaded (green) region).
(b) Three-dimensional (3D) rendering of a vortex structure observed
for the flow of water at $\textnormal{R}\textnormal{e}=75.8$ in a cross-slot with
$\alpha=1.$ The image is generated from $z$-plane images spaced
at $\delta z=5 \mu\textnormal{m}$ and has been
cropped around the central vortex. The volume shown corresponds to
the fluorescently-dyed fluid stream. Also see Movie~M1 for an animated version of Fig.~1(b) \cite{ESI}.}
\label{schematics}
\end{figure}

The cross-slot device is emerging as a promising platform for enhancing mixing of fluids in micro-scale geometries \cite{AitMouheb_etal_2011,AitMouheb_etal_2012}. The planar
elongational flow field generated by the cross-slot geometry (Fig.~1(a)) has found applications in many research areas including for studies of
 macromolecular dynamics \cite{Odell_Keller_1986,Perkins_etal_1997,Schroeder_etal_2003}, extensional rheometry and elastic instabilities of viscoelastic fluids
\cite{Arratia_etal_2006,Poole_etal_2007,Haward_McKinley_2013,Haward_etal_2012}, hydrodynamic trapping \cite{Tanyeri_Schroeder_2013}, and imposing
controlled deformations to complex biological structures \cite{Kantsler_etal_2008,Dylla-Spears_etal_2010,Kantsler_Goldstein_2012,Gossett_etal_2012}.
It has been known since the early 1990's that such intersecting flows are prone to instability beyond a modest critical Reynolds number $\textnormal{Re}_{c}\sim O(10\textrm{--}100)$. In very deep cross-slot flow channels the observed instability takes the form of a stack of three-dimensional vortical structures that appear in the
central cross-over region \cite{Kalashnikov_1991,Kalashnikov_1993}, while in the related 4-roll mill apparatus a flow field incorporating a single helical region has been observed \cite{Lagnado_1990}.  In this work, with cross-slots of modest aspect ratios typical of microfluidic devices (depth to width ratios between 0.4 and 3.87), we have found that the instability occurs above an aspect ratio-dependent value of $\textnormal{R}\textnormal{e}_{c}$ and results in a simple straight vortex that extends downstream along the outlet channels, see Fig.~\ref{schematics}(b). This flow instability has been shown to promote mechanical scission of polymer chains \cite{Vanapalli_2006}, and has more recently been demonstrated to enhance mixing between the two incoming fluid streams in micro-sized devices \cite{AitMouheb_etal_2011,AitMouheb_etal_2012}. Numerical simulations have shown that the mixing quality is higher and $\textnormal{R}\textnormal{e}_{c}$ is lower in the cross-slots than in the more well-known T-shaped mixing device with equivalent channel dimensions \cite{AitMouheb_etal_2012}.

In this Letter, we report the results of detailed experimental and numerical studies of the spiral vortex flow instability in cross-slots with a range of aspect ratios and over a wide range of Re. In contrast to previous studies \cite{Kalashnikov_1991, Kalashnikov_1993, AitMouheb_etal_2012}, we identify appropriate order parameters that characterize the instability as a function of Re in each case. In particular, we present a systematic analysis in terms of bifurcation theory analogous to the Landau theory of phase transitions. The observed phenomena are well described by a Landau-type sixth-order polynomial potential \cite{Landau_1937, Aitta_etal_PRL_1985, Aitta_PRA_1986}, with parameters that show the transition develops from second order to first order as $\alpha$ increases, passing through  a tricritical point for $\alpha \approx 0.55$.  These data can be fully described by scaling theory, and the universal scaling function and behavior are measured near the tricritical point. Improved understanding and characterization of stability conditions for flows through intersecting geometries is vital for the optimization of many laboratory microfluidic experiments and also practical lab-on-a-chip designs, including for the specific goal of enhancing the mixing of fluids in channels with small dimensions operating at low Re.

\begin{figure*}
\includegraphics[width=6.1in]{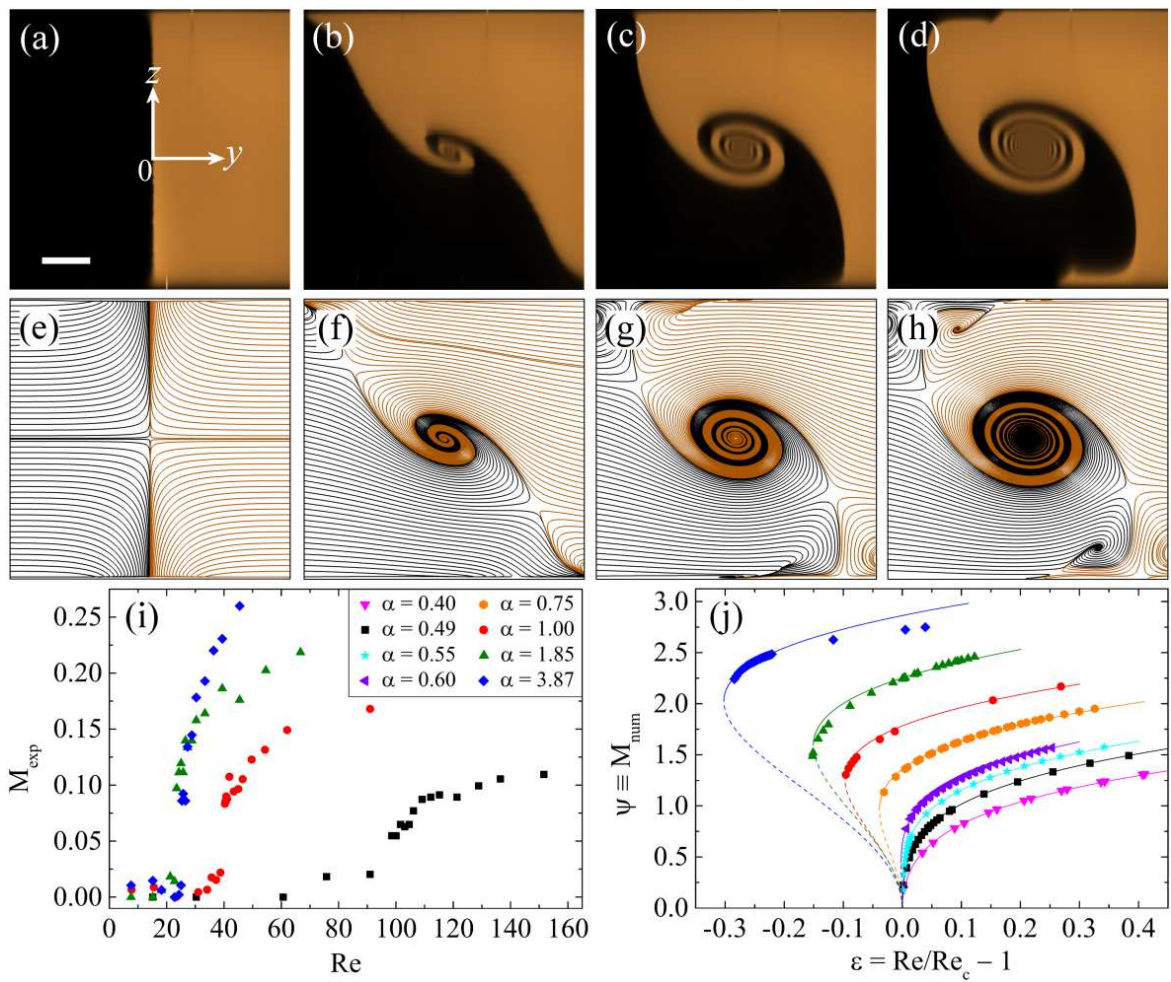}
\caption{(Color online) Confocal imaging (a--d) and numerically generated streamlines (e--h) depicting the evolution of flow structures
in the $x$ = 0 plane for Newtonian fluid flow in the cross-slot device with $\alpha$ = 1 under the following conditions: (a, e) Re = 15.2, (b, f) Re = 42.8, (c, g) Re = 60.6, and (d, h) Re = 91.0. In (a--d) Fluorescently-dyed fluid enters from the right (positive $y$) and undyed fluid enters from the left (negative $y$); outflow is along $x$ (i.e. normal
to the page). Scale bar in (a) represents 200~$\mu$m. In (e--h) the streamlines are colored in order to resemble the experiment. The numerical result shown in (f) is one of two possible solutions at this Re; a symmetric solution can also be obtained, as shown in Fig. S5 \cite{ESI}. (i) Mixing parameter in all four experimental cross-slot devices, evaluated
over the region spanned by $-w/2\leq y\leq w/2$, $-d/2\leq z\leq d/2$ in the $x$ = 0 plane according to Eq.~(1). (j) Numerical order parameter as a function of the control parameter, $\varepsilon$, fitted with a Landau sixth order polynomial potential Eqs.~(3, 4); broken lines represent unstable branches.}
\label{spiral}
\end{figure*}

The cross-slot device (Fig.~\ref{schematics}(a)) consists of two bisecting rectangular
channels of width $w$ and depth $d$, with the aspect ratio defined by
$\alpha=d/w$. In the experiments four devices are utilized, all with
$d=1.2\thinspace\textnormal{mm}$ but with $w$ varied in order to
provide
$\alpha=\textnormal{0.49,\thinspace1.00,\thinspace1.85,\thinspace
and\thinspace3.87}$.  An inlet length of at least 12.5$w$ ensures a 
fully-developed flow before the fluid reaches  the central
region of each device. Newtonian fluid (water) is pumped into two
opposing channels (along the $y$-direction) and exits through the two opposing outlet
channels (along the $x$ direction). One of the incoming fluid streams is
fluorescently-dyed with rhodamine B (concentration = 10~$\mu$M, Sigma) and a
laser-scanning confocal microscope (Zeiss LSM 780) is employed to
examine the interface between fluid streams where they meet in the
central cross-over region. Imaging in
closely-spaced planes through the depth of the flow cell ($-d/2\leq z\leq d/2$) allows
accurate reconstruction of an image in the $x=0$ plane, see (green) shaded area in Fig.~\ref{schematics}(a). Experiments are
performed at 24$^\circ$C  over a range of
$\textnormal{Re}=\rho wU/\mu$, with the fluid density
$\rho=997.1\thinspace\textnormal{kg\thinspace\ensuremath{\mathrm{m}^{\mathnormal{-3}}}}$
and the dynamic viscosity
$\mu=9.1\times10^{-4}\thinspace\textnormal{Pa\thinspace s}$. The
average flow velocity $U$ within each channel of the cross-slot is
controlled using a precision dual syringe pump (Harvard PHD Ultra).
While for low Re the interface between fluid streams is sharp and
vertical over the $y=0$ plane, beyond a fairly moderate critical value
$\textnormal{R}\textnormal{e}_{c}$ the flow bifurcates and breaks symmetry (but remains
steady and laminar) and intricate spiral vortex structures develop, see
Fig.~\ref{schematics}(b) and Movie M1 \cite{ESI}.

The numerical method solves the equations of motion
$\rho\mathrm{D\mathbf{u}}/\mathrm{D}t=-\nabla
p+\mu\nabla^{2}\mathrm{\mathbf{u}}$ and mass conservation
$\nabla\cdotp\mathrm{\mathbf{u}}=0$, for laminar flow of a Newtonian
incompressible fluid by using a fully-implicit, second-order finite volume
method \cite{Poole_etal_2007,Poole_etal_2014,Cruz_etal_2014}. For the
inertial flows considered here, relevant features of the method are the
treatment of the partial time derivative $\partial\mathrm{{\textstyle
\mathbf{u}}}/\partial t$ using the three time-level pressure correction
algorithm \cite{Oliveira_2001} and the treatment of the convective
term $\mathrm{\mathbf{u}}\cdotp\nabla\mathrm{\mathbf{u}}$, which is
discretized using the 3rd order high-resolution CUBISTA scheme
\cite{Alves_etal_2003}. In this way we maintain 2nd-order accuracy in
space and time.

Non-uniform orthogonal meshes are deployed on the 3D cross-slot geometry
of Fig. 1(a), with the central cube (for $\alpha=1$) having $25^{3}$
uniform control volumes (CV's) on the base mesh, and $35^{3}$ or $71^{3}$ CV's
on more refined meshes. The CV size expands slowly towards the two
inlets, along the channel aligned with the $y$-axis, and towards
the two outlets, along the $x$-axis (see Fig.~S1 \cite{ESI}). Theoretical \cite{White} 2D
fully-developed velocity ($v_{th}(x,z)$) profiles are applied at
the inlets, while at the outlets zero axial gradients are assumed
for velocity components ($\partial/\partial x=0$), pressure is linearly extrapolated and the total
flow rate is forced to satisfy overall mass conservation ($Q_{out}=2Q_{in}$,
where $Q_{in}=Uwd$ for each inlet section). Note that we do not force
the flow rate entering via each inlet to divide equally between
each outlet, but only that the flow rate through each outlet is equal, as in the experiments. The numerical simulations explore devices with aspect ratios set equal
to the four experimental devices as well as additional values of aspect
ratio near to the tricritical value.

Great care is exercised to guarantee that the numerical results are
well converged and do not significantly with the computational mesh. In Fig. S1 \cite{ESI}, we illustrate the various meshes employed and compare the results obtained for a cross-slot with $\alpha=1$, showing good accuracy.

In Fig.~\ref{spiral}(a--d) we present reconstructed confocal images of the $x=0$
plane (spanning $-w/2\leq y\leq w/2$, $-d/2\leq z\leq d/2$) for
the cross-slot device with $\alpha$ = 1, which show the evolution
of the vortical structure as Re increases. Qualitatively similar behavior
is observed in all four geometries (see Fig.~S2 and Movies M2--M5 \cite{ESI}): at low Re the interface between fluid streams is sharp
and vertical as expected, however as Re increases above a critical value the spiral vortex forms abruptly about the $x$-axis. With further
increases in Re, the spatial extent of the spiral expands, as does the number of
turns of the arms. The value of the critical Reynolds
number $\textnormal{R}\textnormal{e}_{c,exp}$ for the appearance of the
spiral in the experiments depends on the aspect ratio, but is broadly consistent with
values reported previously in related experiments \cite{Kalashnikov_1991,Kalashnikov_1993,AitMouheb_etal_2012,Lagnado_1990} ($\textnormal{Re}_{c,exp} \approx$ 100, 40, 24 and 26 for $\alpha =$ 0.49, 1.00, 1.85 and 3.87, respectively).
In each of the four devices the central vortex forms with a favored
orientation about the $x$-axis (anticlockwise for $\alpha=0.49$
and $\alpha=1.00$ and clockwise for $\alpha=1.85$ and $\alpha=3.87$,
see Fig.~S2). The orientation is presumably biased by some minor geometrical
imperfections in the devices and causes the bifurcation to follow
a favored branch. In only a few instances in the cross-slots with
$\alpha=1.85$ and $\alpha=3.87$ at higher Re did we observe the vortex
to form in the unfavored sense (see Fig.~S3 \cite{ESI} for examples).
The vortex structures observed in the $x=0$ plane can be well-fitted
by elliptical logarithmic spirals (see Fig.~S4 \cite{ESI}).

Numerical simulations result in remarkably good agreement with the
experiment, as shown by the streamline plots in Fig.~\ref{spiral}(e--h) that can be directly compared with the experimental results in Fig.~\ref{spiral}(a--d). Note that in the numerical simulations at $\alpha=1.00$ there is hysteresis in the transition and both symmetric and asymmetric steady solutions can be obtained for $\textnormal{R}\textnormal{e}=42.8$ (Fig.~\ref{spiral}(f)), depending on whether Re is quasistatically increased or decreased from below or above the transition, respectively. Both possible solutions are presented in Fig.~S5 \cite{ESI}. Additional numerical plots of velocity vector fields at various Re spanning the transition for the case of $\alpha=1.00$ are provided in Fig.~S6 \cite{ESI}. Our  experimental method is not amenable to a genuinely quasistatic variation of Re since the flow is necessarily interrupted between image acquisition at each Re in order to refill the syringes with fluid. Therefore our experiment can not resolve the hysteresis in the transition.

We treat our experimental data following the approach suggested by
Stroock et al.~\cite{Stroock_etal_2002} in order to assess the degree of
mixing between the incoming fluid streams, $M_{exp}$. This parameter is computed
using the standard deviation of the pixel intensity evaluated over
images such as those shown in Fig.~\ref{spiral}(a--d) and Fig.~S2 \cite{ESI}:
\begin{equation}
M_{exp}=1-2\langle (I- \langle I \rangle)^2 \rangle^{1/2},
\end{equation}
where $I$ is the grayscale pixel value (normalized between 0 and
1, corresponding to the minimum and maximum grayscale intensities in the images) and $\langle \rangle$ indicates an average over all pixels in the field of
view. For completely segregated fluid streams, there is a binary distribution of pixel intensities, the standard deviation $=0.5$ and hence $M_{exp}=0$; this
is essentially the case illustrated in Fig.~\ref{spiral}(a). Complete mixing between the fluid
streams would mean a uniform pixel intensity of $I=0.5$ over the
entire field of view, therefore a standard deviation of zero and hence $M_{exp}=1.$ In Fig.~\ref{spiral}(i), we plot the mixing
parameter, $M_{exp}$, as a function of the Reynolds number for all
four experimental aspect ratios. Fig.~\ref{spiral}(i)  shows a general
reduction in $\mathrm{R}\mathrm{e}_{c,exp}$ and an increasingly
large and rapid increase in $M_{exp}$ as $\alpha$ increases. We
also note that the transition is imperfect; there is a small but distinct curvature
in $M_{exp}$ near the onset, which is consistent with the presence
of their being a favored branch for the transition (clockwise or anticlockwise spirals).
We treat the favored branch as the positive branch regardless of the
spiral orientation.

In the case of the numerical simulations, we find the growth of the instability
is best described by the following parameter:
\begin{equation}
M_{num}=\frac{\left.v_{max}\right|_{x=y=0}}{U},
\end{equation}
where $\left.v_{max}\right|_{x=y=0}$ is the maximum value of the
$y$-component of the velocity measured along the $z$-axis at each Re (for a symmetric flow $M_{num} \equiv 0$). A pictorial description of $M_{num}$ is given in Fig.~S6 \cite{ESI}. Fig.~\ref{spiral}(j) shows a plot of $M_{num}$ as a function of the control parameter $\varepsilon= (\text{Re}-\text{Re}_c)/\text{Re}_c$ for various aspect ratio cross-slot devices. The data in Fig.~\ref{spiral}(j) is
fitted using a Landau-type model with "free energy" $F$, given by:
\begin{equation}
F=-h\psi- \frac{1}{2}\varepsilon \psi^{2}+\frac{g}{4}\psi^{4}+\frac{k}{6}\psi^{6},
\label{eqF}
\end{equation}
where the order parameter $\psi \equiv M_{num}$, in the numerical case. Even terms are included in Eq. (3) since for a perfect system $F$
should be independent of the sign of $\psi$ (i.e. the handedness of
the spiral). The lowest order asymmetric ``field" term ($h\psi$) can account for imperfections that bias the handedness. In the numerical simulations the vortex can form in the clockwise
or anticlockwise orientation with similar probablity, hence $h=0$ (for
all aspect ratios). For given values of the parameters $h$, $g$, and $k$, the
value of $\psi(\varepsilon)$ corresponds to the extrema of $F$,  $\partial F /
\partial \psi = 0$, giving:

\begin{equation}
\varepsilon \equiv \text{Re} / \text{Re}_c - 1 = k \psi^4 + g\psi^2 - h\psi^{-1}. \label{eqM}
\end{equation}

The ratio of the coefficients $g/k>0$ for $\alpha\lesssim0.55$
(corresponding to a forward bifurcation or a second-order transition)
and decreases monotonically with increasing $\alpha$, turning negative
for $\alpha\gtrsim0.55$ at which point the bifurcation becomes backwards,
or first order and an increasing large hysteresis loop grows as $\alpha$ increases. The numerical data obtained for $\alpha=0.55$ are well-fitted by Eq.~(4) with $g=0$, (cyan) stars in  Fig.~\ref{spiral}(j), and thus corresponds to a tricritical transition. Fitting of the numerical data with Eq.~(4) provides the values of $\mathrm{R}\mathrm{e}_{c}$ shown
in Fig.~S7 \cite{ESI}. For $\alpha\gtrsim0.55$, due to the hysteresis
loop we can find two values of the critical Reynolds number: $\mathrm{R}\mathrm{e}_{c}$ denotes the value found for quasistatic increases
in Re from below, while $\mathrm{R}\mathrm{e}_{c}^{*}$ denotes
the value found for quasistatic decreases in Re from above the onset.
The numerically predicted hysteresis loops for $\alpha\gtrsim0.55$ can not be observed experimentally, however we find that our experimental critical Reynolds numbers ($\mathrm{R}\mathrm{e}_{c,exp}$) agree very well (within approx.  $\pm 5 \%$) with the numerical values of $\mathrm{R}\mathrm{e}_{c}^{*}$ (Fig.~S7).

\begin{figure}
\includegraphics[width=2.9in]{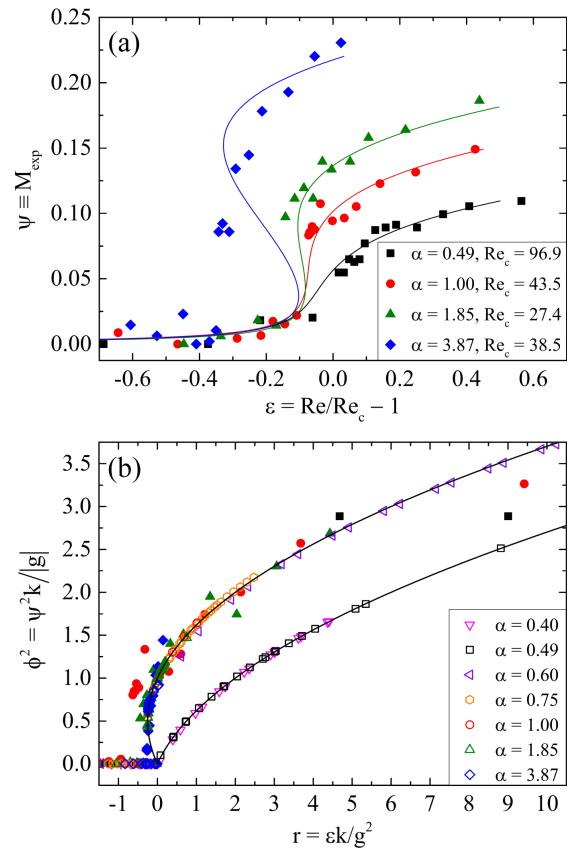}
\caption{(Color online) (a) Experimental mixing parameter fitted with the sextic Landau potential indicated by the numerical results. (b) Collapse
of experimental data (solid symbols) and numerical data (hollow symbols) onto the theoretical curve given by Eq.~(5) (solid lines).}
\label{scaling}
\end{figure}

In Fig.~\ref{scaling}(a) we fit our experimental mixing parameter with
Eq.~(4), setting the order parameter $\psi \equiv M_{exp}$ and fixing $\text{Re}_c$ to the value determined
numerically for each aspect ratio device. We introduce a small
coefficient $0.002< h < 0.003$ into Eq.~(4) that accounts for geometrical
imperfections in the experimental devices. In general we find that our experimental
order parameter is well described by the same sextic Landau potential
that describes the numerical data.

The theory for tricritical points \cite{goldenfeld1992lectures} predicts a universal scaling form for
the canonical order parameter $\phi \equiv \psi \sqrt{k/|g|}$ in terms of
the control parameter $r\equiv \varepsilon k/g^2$, showing data collapse in the form:
\begin{equation}
\phi^2 = (1 - \text{sgn}(g)\sqrt{1+4r})/2.
\label{scalingfn}
\end{equation}
In Fig.~\ref{scaling}(b), we have plotted both the experimental and numerical results
in scaled form and show the comparison with the data collapse
prediction.  The agreement is excellent and confirms our identification
of the mixing transition as a tricritical one.

In summary, we have demonstrated that the spiral instability observed
for Newtonian flow in the cross-slot device driven far from equilibrium above a critical Reynolds number, can be well-described by a Landau
model analogous to that used near equilibrium tricritical points.
We note a certain similarity of our results with those presented for
flow in low-aspect ratio Taylor-Couette devices above a critical angular
velocity (e.g. \cite{Aitta_etal_PRL_1985,Aitta_PRA_1986,Benjamin_etal_1981}),
however the instability in the cross-slot geometry can be harnessed
to promote mixing of Newtonian fluids at the quite modest Reynolds numbers accessible in  microfluidic devices.
Such improved mixing could benefit the efficiency of lab-on-a-chip applications such as chemical synthesis, antibody antigen binding reactions and bioassay.  It is likely that bifurcation phenomena reported in related flow geometries (such as in the T-shaped micromixer, for example  \cite{AitMouheb_etal_2012, Poole_etal_2013}) could also be characterized by the tricritical point formalism presented here. 

\vspace{5pt}

\noindent{\bf Acknowledgements:}  SJH and AQS gratefully acknowledge the support of the Micro/Bio/Nanofluidics Unit of the Okinawa Institute of Science and Technology Graduate University. MAA acknowledges financial support from the European Research Council (Grant Agreement No. 307 499).
We are indebted to Professors Helmut Brand and Gregory Falkovich for helpful discussions. 

\bibliographystyle{apsrev4-1}
\bibliography{xampl,x-slot_Newt}

\end{document}